\documentstyle[thmsa,11pt,a4,sw20lart]{article}

\input tcilatex
\begin{document}

\topmargin 0pt \oddsidemargin 5mm

\setcounter{page}{1}

\hspace{8cm}{} \vspace{2cm}

\begin{center}
{\large {STOPPING OF GYRATORY FAST PARTICLE IN MAGNETIZED COLD PLASM}}\\

$^{(1)}${H.B. Nersisyan, }$^{(1)}$A.V. Hovhannisyan and $^{(2)}$C. Deutsch\\%
\vspace{1cm} $^{(1)}${\em Division of Theoretical Physics, Institute of
Radiophysics and Electronics, 2 Alikhanian Brothers St., Ashtarak-2, 378410,
Republic of Armenia}\footnote{%
E-mail: Hrachya@irphe.sci.am}\\

$^{(2)}${\em Laboratoire de Physique des Gaz et Plasmas, B\^at.10,
Universit\'e Paris XI, 91405 Orsay, France}
\end{center}

\vspace {5mm} \centerline{{\bf{Abstract}}}

The energy loss by a test gyratory particle in a cold plasma in the presence
of homogeneous magnetic field is considered. Analytical and numerical
results for the rate of energy loss are presented. This is done for strong
and weak fields (i. e., when the electron cyclotron frequency is either
higher, or smaller than the plasma frequency), and in case, when the test
particle velocity is greater than the electron thermal velocity. It is shown
that the rate of energy loss may be much higher than in the case of motion
through plasma in the absence of magnetic field.

\newpage 

The energy loss of fast charged particles in a plasma has been a topic of
great interest since the 1950s because of its considerable importance for
the study of the basic interactions of charged particles in real media;
moreover, recently, it has also become a great concern in connection with
heavy-ion driven inertial fusion research [1, 2].

The nature of experimental plasma physics is such that experiments are
usually performed in the presence of magnetic fields, and consequently, it
is of interest to investigate the effects of a magnetic field on the rate of
energy loss.

The stopping of charged particles in a magnetized plasma has been the
subject of several papers [3-6]. The stopping of a fast test particle moving
with a velocity $v$ much higher than the electron thermal velocity $v_T$ was
studied in refs. [3, 5]. The energy loss of a charged particle moving with
arbitrary velocity was studied in [4]. The expression derived there for the
Coulomb logarithm corresponds to the classical description of collisions.

In ref. [6] expressions were derived describing the stopping power of a slow
charged particle in Maxwellian plasma with a classically strong (but not
quantizing) magnetic field, under conditions such that the scattering
processes must be described quantum-mechanically.

In the present paper we consider the rate of energy loss of a
nonrelativistic gyratory charged particle in a magnetized cold plasma. Also,
this problem is important for the construction of models of $X$-ray pulsars
[7] and the study of processes in the atmospheres of magnetic white dwarfs
the magnetic fields on the surfaces of which can attain strengths of $%
10^5-10^{10}kG$.

A uniform plasma is considered in the presence of a homogeneous magnetic
field ${\bf B}_0$ (directed in the positive $z$-direction) which is assumed
sufficiently small so that $\lambda _B\ll a_c$ (where $\lambda _B$ and $a_c$
are respectively the electron de Broglie wavelength and Larmor radius). From
these conditions we can obtain $B_0<10^5T$ ($T$ is the plasma temperature),
where $T$ is measured in $eV$ and $B_0$ in $kG$. Also, due to the high
frequencies involved, the very weak response of the plasma ions is neglected
and the Vlasov-Poisson equations to be solved for the perturbation to the
electron distribution function and the scalar potential $\varphi $. The
solution is (see, for example, [8])

\begin{equation}
\varphi ({\bf r},t)=\frac{4\pi Ze}{(2\pi )^4}\int d{\bf k}\int_{-\infty
}^{+\infty }d\omega \frac{\exp [i({\bf kr}-\omega t)]}{k^2\varepsilon ({\bf k%
},\omega )}\int_{-\infty }^{+\infty }d\tau \exp [i\omega \tau -i{\bf kr}%
_0(\tau )],
\end{equation}
where ${\bf r}_0(t)$ is the radius-vector of the test particle having the
components $x_0(t)=a\sin (\Omega _ct)$, $y_0(t)=a\cos (\Omega _ct)$, $%
z_0(t)=0$ ($\Omega _c=ZeB_0/Mc$, $a=v/\Omega _c$, $Ze$ and $v$ are the
Larmor frequency, the Larmor radius, the charge and the velocity of the test
particle respectively), $\varepsilon ({\bf k},\omega )$ is the longitudinal
dielectric function of magnetized cold plasma

\begin{equation}
\varepsilon ({\bf k},\omega )=\varepsilon (\omega )\cos ^2\alpha +h(\omega
)\sin ^2\alpha
\end{equation}
with

\begin{equation}
\varepsilon (\omega )=1-\frac{\omega _p^2}{\omega (\omega +i\upsilon )},
\end{equation}

\begin{equation}
h(\omega )=1+\frac{\omega _p^2(\omega +i\upsilon )}{\omega \left[ \omega
_c^2-(\omega +i\upsilon )^2\right] }.
\end{equation}
Here, $\alpha $ is the angle between the wave vector ${\bf k}$ and the
magnetic field, $\omega _p=\sqrt{4\pi n_0e^2/m}$, $\omega _c$ and $\upsilon $
are the plasma frequency, Larmor frequency and the effective collision
frequency of the plasma electrons respectively.

The rate of energy loss $S$ of a plasma against a fast charge is defined by
the energy loss of the charge in a unit time due to interactions with the
plasma electrons. From eq. (1) it is straightforward to calculate the
electric field ${\bf E}=-{\bf \nabla }\varphi $, and the stopping force
acting on the particle. Then, the rate of energy loss of the test particle
becomes

\begin{equation}
S=\frac{2Z^2e^2\Omega _c^2}{\pi v}\sum_{n=1}^\infty nQ_n(s){\rm Im}\left[ 
\frac{-1}{\varepsilon (n\Omega _c)T(n\Omega _c)}\right] ,
\end{equation}
where

\begin{equation}
T(\omega )=\sqrt{\frac{\left| P(\omega )\right| +{\rm Re}P(\omega )}2}+i{\rm %
sgn}\left[ {\rm Im}P(\omega )\right] \sqrt{\frac{\left| P(\omega )\right| -%
{\rm Re}P(\omega )}2},
\end{equation}

\begin{equation}
Q_n(s)=\pi \int_0^sdxJ_n^2(x),
\end{equation}
$P(\omega )=h(\omega )/\varepsilon (\omega )$, $J_n(x)$ is the $n$th order
Bessel function and $s=k_{\max }a$ with $k_{\max }=1/r_{\min }=2mv/\hbar $,
where $r_{\min }$ is the effective minimum impact parameter. Here $k_{\max }$
has been introduced to avoid the divergence of the integral caused by the
incorrect treatment of the short-range interactions between the test
particle and the plasma electrons within the linearized Vlasov theory.

The function $Q_n(s)$ is exponentially small at $n>s$. Therefore the series
in the eq. (5) is cut at $n_{\max }\simeq s$ and the rate of energy loss is
determined by harmonics with $n<n_{\max }$.

Consider now the eq. (5) for strong and weak magnetic fields. In the case of
weak magnetic field ($\Omega _c<\upsilon $) one may substitute the summation
in expression (5) by integration in $\omega =n\Omega _c$. Since also $%
Q_n(s)\simeq \ln (s/n)$ when $s>n$, in the limit $\Omega _c\rightarrow 0$
the eq. (5) is transformed into a known Bohr's expression.

Consider the case of strong magnetic field and let be non-integer. In this
case, from eq. (5) we find:

\begin{equation}
S\simeq \frac{Z^2e^2\omega _p^2}{\pi v}\frac \upsilon {\Omega
_c}\sum_{n=1}^\infty \frac 1{n^2}Q_n(s)\left[ 1+\frac{n^4}{\left(
n^2-c^2\right) ^2}\right] .
\end{equation}
From eq. (8) it follows, that energy loss decreases inversely proportional
to the magnetic field. In the case when $c=1$ (electron test particle), from
eq. (5) we find:

\begin{equation}
S\simeq \frac{Z^2e^2\omega _p^2}{\pi v}\frac{\Omega _c}\upsilon Q_1(s).
\end{equation}
Note that the rate of energy loss increases proportionally to the magnetic
field.

These examples of asymptotic dependence of energy loss rate on the value of
magnetic field show strong dependence of energy loss on mass of the test
particle.

From the eq. (5) it is straightforward to trace qualitatively the behavior
of energy loss rate as a function of magnetic field in the general case.
Note, as it follows from eq. (5) that the rate of energy loss is maximal for
those values of magnetic field for which $\varepsilon (n\Omega _c)$ has
small values. The small $\varepsilon (n\Omega _c)$ means, that in the
dependence of energy loss from magnetic fields, maximums at integer values
of parameter $b=\omega _p/\Omega _c$ can be observed. It corresponds to the
case, when on test particle's Larmor orbit includes integer number of plasma
oscillation wavelengths ($\lambda _p=2\pi v/\omega _p$).

Fig.1 shows the ratio $R=S/S_B$ (where $S_B$ is the well-known Bohr result
[8]) as a function of parameter $b$ in two cases; for electron test particle
(dotted line), and for proton test particle (solid line). The
plasma-particle parameters are: $\upsilon /\omega _p=0.01$, $%
n_0=10^{18}cm^{-3}$, $T=100eV$ and $v/v_T=10$. As it follows from Fig.1, the
rate of energy loss oscillates as a function of magnetic field and many
times exceeds the usual Bohr losses of energy.

\begin{center}
\newpage\ Figure Caption
\end{center}

Fig.1. The ratio $R=S/S_B$ as a function of parameter $b$ in two cases; for
electron test particle (solid line), and for proton test particle (dotted
line). The parameters are: $\upsilon /\omega _p=0.01$, $n_0=10^{18}cm^{-3}$, 
$T=100eV$ and $v/v_T=10$.

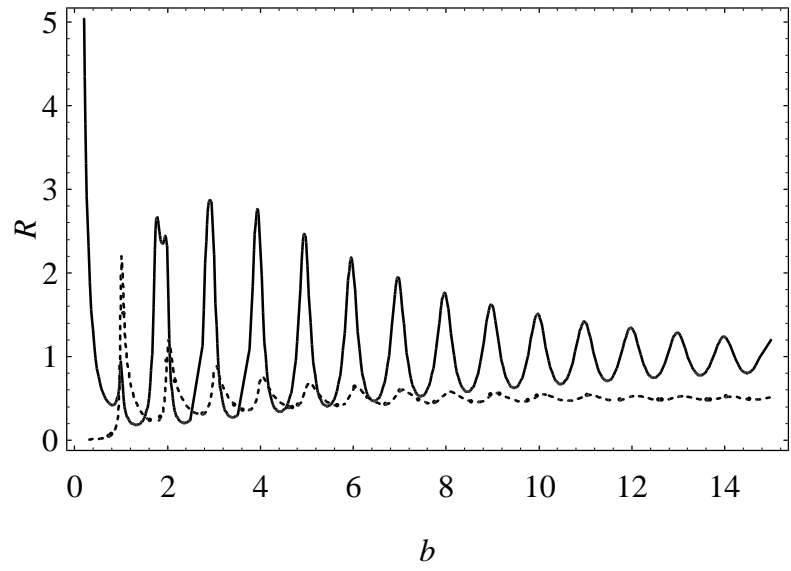
\end{document}